# Percolation Effect Induced Significant Change of Complex Permittivity and Permeability for Silver-Epoxy Nano-Composites

**Bo-Wei Tseng and Tsun-Hsu Chang** [a)]

## AFFILIATIONS

Department of Physics, National Tsing Hua University, 30013, Hsinchu, Taiwan.

[a)] **Author to whom correspondence should be addressed:** thschang@phys.nthu.edu.tw

**ABSTRACT** **(~ 250 words)**

The intricate interplay between complex permittivity and permeability constitutes the cornerstone of electromagnetic (EM) applications, enabling precise customization for various uses. This study employed silver-epoxy nano-composites to exemplify a conductor-insulator composite, leveraging silver's exceptional attributes, such as high conductivity and low reactivity. The determination of complex permittivity and permeability was conducted via the transmission/reflection method. At lower concentrations of dispersed silver particles, these nano-particles within the epoxy resin act as modest dipoles, augmenting permittivity. This regime aligns closely with the effective medium theory (EMT) and comprises the focus of much research. However, nearing the percolation threshold, a percolation effect emerges, drastically accelerating enhancement rates beyond the predictions of EMT. Simultaneously, long-wavelength electromagnetic waves induce diamagnetic currents within loops formed by metal grains. This diamagnetic effect intensifies with increasing volume fraction, leading to a reduction in permeability. This study observed percolation power law behavior near the threshold with calculated critical exponents. Consequently, the dielectric constant of the silver-epoxy nano-composite reached a maximum of 515. Regarding the relative permeability, the lowest recorded value was 0.31. These findings were obtained within the X-band (8.2 GHz~12.4 GHz) region

Metamaterials and composites have grabbed attention in recent years since they can possess unique properties that don't exist in nature. Complex permittivity ($\varepsilon_0\varepsilon' + i\,\varepsilon_0\varepsilon''$) and permeability ($\mu_0\mu' + i\mu_0\mu''$) both characterize metamaterials. The ability to manipulate these properties provides an extraordinary opportunity to tailor materials according to specific requirements, enabling a wide range of applications using a single material. High-k composites with low loss find extensive use in microwave devices [1-5], while high-loss and negative susceptibility metamaterials have applications in stealth technology. Furthermore, metamaterials have diverse applications in antenna design [6, 7], split-ring resonators [8], medical imaging [9, 10], wireless power transfer [11-14], anti-reflection coating [15, 16], optical devices [17, 18], and negative refractive index materials [19-23].

Effective medium theory (EMT) is a widely adopted approach to analyzing composites' material properties, enabling fine-tuning permittivity and permeability. However, its efficacy diminishes when the volume fraction approaches the percolation vol. fraction threshold. For instance, Yao *et al.* have collected the ten most representative EMT models [24], all of which can only handle a small region in volume fraction, and none of those predicted a threshold. Also, EMT is limited in $\varepsilon'$. EMT cannot explain $\varepsilon''$, $\mu'$ and $\mu''$. This is where the percolation theory kicks in.

To handle EMT's drawback, percolation theory is then applied, which was first introduced in 1957 by Boardbent and Hammersley [25]. Yet, most investigations in the last six decades have focused on the dielectric constant ($\varepsilon'$) and the electrical conductivity ($\sigma$). Roughly speaking, research of $\varepsilon''$ [26] and $\mu' + i\mu''$ [27-29] stagnated after the 1990s due to the lack of profound application. There seems to be no room for further exploration, but the necessity of understanding ($\varepsilon$, $\mu$) still serves scientific interests (etc., refraction index n($\varepsilon$, $\mu$) and wave impedance, Z($\varepsilon$, $\mu$)). In general, if vol. fraction p is far away from the percolation threshold, $p_c$, it is safe to suppose $\mu = 1$ until the vol. fraction approaches $p_c$.

Permeability contribution (in terms of $\mu(p - p_c)$) seems inevitable when $p \to p_c$. In this situation, the assumption $\mu \doteq 1$ should be slightly modified. Su *et al.* showed that silver-epoxy composite exhibits diamagnetism while epoxy and silver are both non-magnetism [30]. Lagar'kov *et al.* showed that the permeability of a ferromagnetic percolation system also has diamagnetic percolative behavior [29]. Bowman and Stroud showed that superconductor-metal composites also possess diamagnetism and follow the percolation power law [28]. They also predicted that metal-insulator composites also have similar properties. Therefore, magnetism may not be rare in composites, and if permeability is assigned to be 1, then the permittivity could be over/underestimated.

Now, in formulae, the behavior of a composite's dielectric constant can be stated as [31, 32]

$$\varepsilon' = \begin{cases} \varepsilon_d'|p - p_c|^{-s} & p < p_c, p > p_c \quad (1a) \\ \varepsilon_d'^{\frac{t}{t+s}}\left(\frac{\sigma_m}{\omega}\right)^{\frac{s}{t+s}} & p \text{ near } p_c \quad (1b) \end{cases}$$

$p$ is the volume fraction, $p_c$ is the percolation threshold; $\varepsilon_d'$ is the dielectric constant of the insulator; $\sigma_m$ is the conductivity of the metal, $s$ and $t$ are the critical exponents. The theoretical value is 0.7~1.0 for $s$ in 3D. Near the threshold, there is a remarkable surge in the dielectric constant, reaching a tremendously significant value. However, this substantial enhancement is confined to a narrow vicinity. Consequently, accurately calculating the volume fraction near the threshold poses a challenging task. The imaginary part also follows the power law [26]

$$\varepsilon'' \propto \begin{cases} |p-p_c|^{-q} & p < p_c \quad (2a) \\ \left(\frac{\sigma_m}{\omega}\right)^{\frac{s}{t+s}} & p \text{ near } p_c \quad (2b) \end{cases}$$

Theoretically, the critical exponent $q$ is about 2.5~3.1 in 3D.

According to Bowman's model, long-wavelength electromagnetic waves induce diamagnetic currents in loops formed by metal grains. This diamagnetic effect intensifies with an increase in volume fraction, causing a permeability drop due to the more fabulous presence of "bond" electrons. However, as the volume fraction continues to rise, the swift expansion of silver nano-particle clusters prompts the 'bond' electrons to transition back to a 'free' state. Consequently, the induced current weakens, leading to an increase in permeability. Then, the "negative magnetic susceptibility" $(-\chi_b)$ follows the power law near the threshold [27, 28]

$$-\chi_b \propto |p-p_c|^{-u} \qquad p \text{ close to } p_c \qquad (3)$$

The critical exponent $u$ for the negative magnetic susceptibility has a theoretical value of 0.11~0.41 [28]. As for the imaginary part, it reads [28, 29]

$$\mu'' \propto |p-p_c|^{-v} \qquad p \text{ close to } p_c \qquad (4)$$

According to the theory, critical exponent, $v$ , is equal to $u$ [28].

Our sample preparation closely followed the established waveguide method [33]. The waveguide cleaning and fabrication process of the composite is visually depicted in Fig. 1. Commencing with a WR90 copper waveguide, we initiated the process by immersing it in a copper polishing solution to eliminate surface oxidation. Particular attention was given to the interior surface to prevent conducting losses, ensuring that the losses were solely attributed to the composite material. This initial polishing step, lasting approximately 10 to 20 seconds, allowed ample time for the solution to react with the oxidation.

Subsequently, the waveguide underwent immersion in distilled water to eliminate the residual polishing solution. A thorough cleansing process ensued, involving washing with acetone to remove water and any organic substances from the surface. The final step in the washing process involved an isopropanol bath to eliminate traces of acetone. With the washing process complete, the waveguide was ready to be affixed to the Teflon holder.

The holder comprised five components: a metal base, two WR90-shaped Teflon plates, and two Teflon plates. Structurally resembling a sandwich, the holder featured the metal base at the bottom, followed by a plate, with the WR90-shaped plates flanking the copper waveguide. Before sealing the waveguide, the composite mixture was introduced.

The nano-powder utilized in this study was commercially sourced from US Research Nanomaterials, Inc., boasting a purity exceeding 99.99%, a diameter of 20 nm, and a density of 10.49 g/cm³. The epoxy resin (Epoxy A) employed possessed notable adhesive properties. When combined with the hardener (Epoxy B), the epoxy exhibited a density of about 1.13 g/cm³.

The fabrication began by pouring one to three grams of Epoxy A into a flask, then heated to approximately 65 ℃ on a heating plate. This temperature adjustment aimed to reduce viscosity, facilitating a more accessible and random distribution of the silver nano-powder. Subsequently, the desired volume fraction of powder was added, and a moderate stirring ensued. Once the mixture achieved homogeneity, Epoxy B was introduced. The stirring and heating process continued until the liquidity of the mixture significantly diminished. The resulting composite was then poured into the Teflon holder, sealed with the final round plate, and secured with screws. The assembly was prepared for the curing process.

The holder was placed in a custom-made oven capable of reaching temperatures around 80 ℃ to expedite curing. The metal base, designed to be affixed to a motor inside the oven, provided the sample with a moderate rotation during curing, preventing nano-particle precipitation. While the curing process under room temperature would take approximately 20 hours, elevating the temperature to 70-90 ℃ allowed the composite to harden within 3 hours.

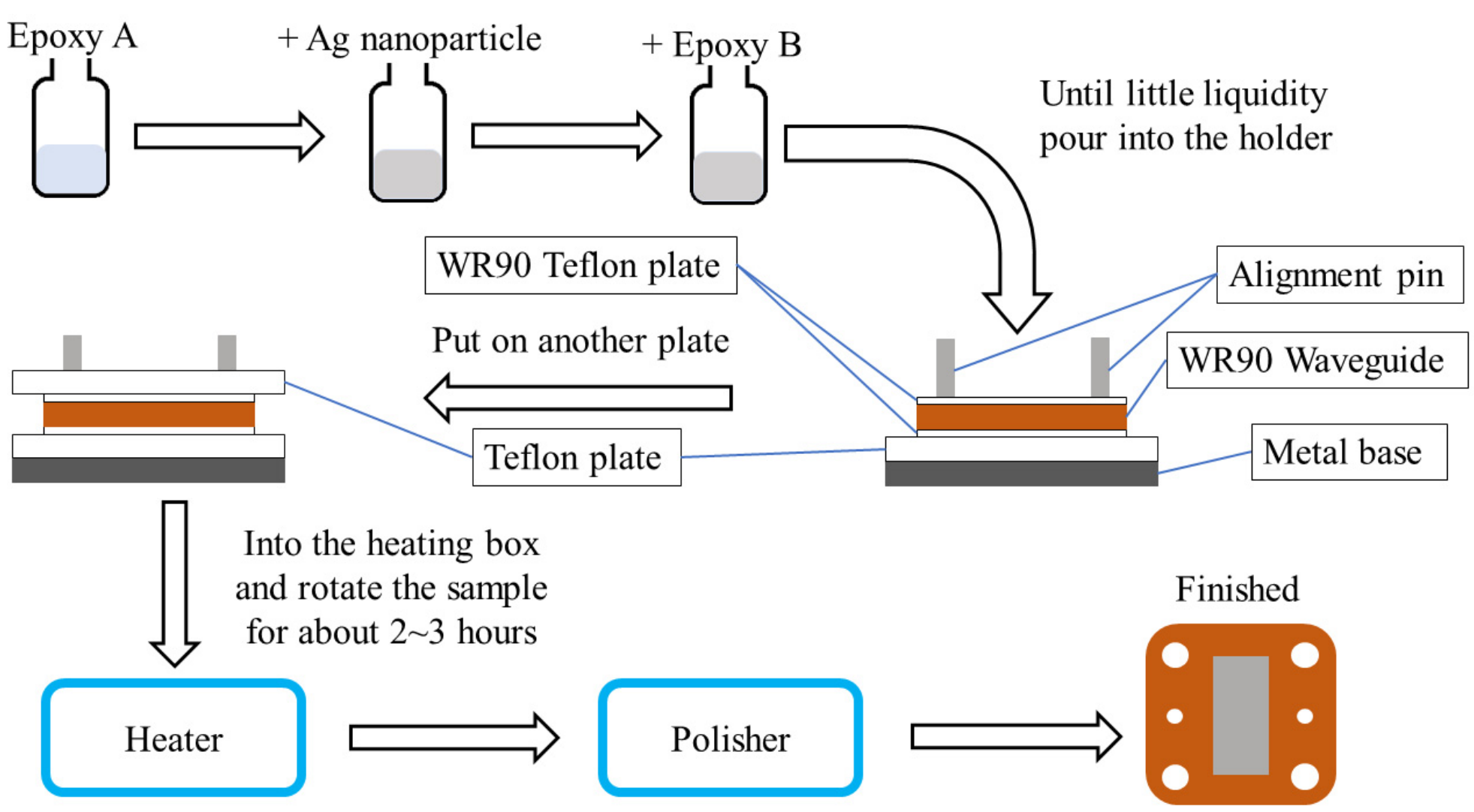

**FIG. 1.** The flow chart of the sample preparation for the silver-epoxy composites with a WR-90 standard rectangular waveguide.

Upon completion of curing, the composite underwent a polishing phase. Starting with 500-grit sandpaper, excess material from the waveguide was removed, followed by successive steps with 1000, 1500, 2500, and finally, 4000-grit sandpaper. The desired outcome was a mirror-like shine for the waveguide and a smooth surface for the composite, devoid of any cavities, pores, or holes to ensure the acquisition of precise results. With the sample now prepared, it was ready for measurement. The waveguide cleaning and the fabrication process of the composite are illustrated in Fig. 1.

The measurements were conducted using the Keysight PNA-X N5247B instrument. Two coaxial cables were connected to its two ports, and two adapters, transforming coaxial to WR90 waveguide, were affixed to the cables, extending the operational bandwidth from 8.2 GHz to 12.4 GHz (X-band). The calibration kit employed was the X11644A, designed explicitly for the measurements. All measurements were carried out under room temperature conditions (around 300 K).

The sample, positioned between the two ports, underwent S-parameter measurements. Subsequently, the acquired data was subjected to analysis through the Fresnel equation and the infinite geometric sequence. MATLAB (R2021b) was utilized for the computation, allowing the retrieval of permittivity and permeability from the S-parameters.

The Fresnel equation reads,

$$r=\frac{Z_0-Z}{Z_0+Z} \quad r'=\frac{Z-Z_0}{Z_0+Z}$$
$$t=\frac{2Z}{Z_0+Z} \quad t'=\frac{2Z_0}{Z_0+Z} \tag{5}$$

where $Z_0$ is the impedance of vacuum in the waveguide, and $Z$ is the impedance of the composite in the waveguide. The signal measured is the combination of multiple reflections and transmissions so that the infinite geometric sequence can be written down.

$$\begin{cases} S_{11}=r+\dfrac{tr't'e^{2i\beta d}}{1-r'^2e^{2i\beta d}} \\ S_{21}=\dfrac{tt'e^{2i\beta d}}{1-r'^2e^{2i\beta d}} \end{cases} \tag{6}$$

where $\beta$ is the wave number of the composite inside the waveguide of $TE_{01}$ mode.

Fig. 2. illustrates the frequency dependence of complex permittivity and permeability for samples with various volume fractions. When the volume fractions are below 25% in Fig. 2(a), the frequency responses across the whole bandwidth are weak. However, when the volume fractions exceed 25%, e.g., 25.01% and 27.04%, the frequency dependence becomes visible. Figs. 2(b), 2(c), and 2(d) also have a similar trend that when the volume fraction is close to the percolation threshold, frequency response becomes apparent, as predicted in Eq. 1(b). We will discuss this issue in more detail later.

Fig. 3(a). is the result of $\varepsilon'$ taken at 8.2 GHz (blue triangles), and 12.4 GHz (red dots). It starts from pure epoxy with a dielectric constant of around 3.13. $\varepsilon'$ gradually increases from 3.2 to 8.6 as the volume fraction rises from 0% to 10%. Within this range, the silver nano-particles are isolated and can be treated as dielectric material, allowing for the application of EMT [30]. A more pronounced increase occurs at 15% to 24%, and as we further increase the volume fraction, the rate of increasement (the slope) surge swiftly, and slows down when the volume fraction exceeds 25%, where the maximum relative permittivities of 515 (8.2 GHz) and 322 (12.4 GHz) were reached.

The fitting results are given in Fig. 3(b) on a logarithmic scale. The percolation thresholds are 26.25% and 26.34%,

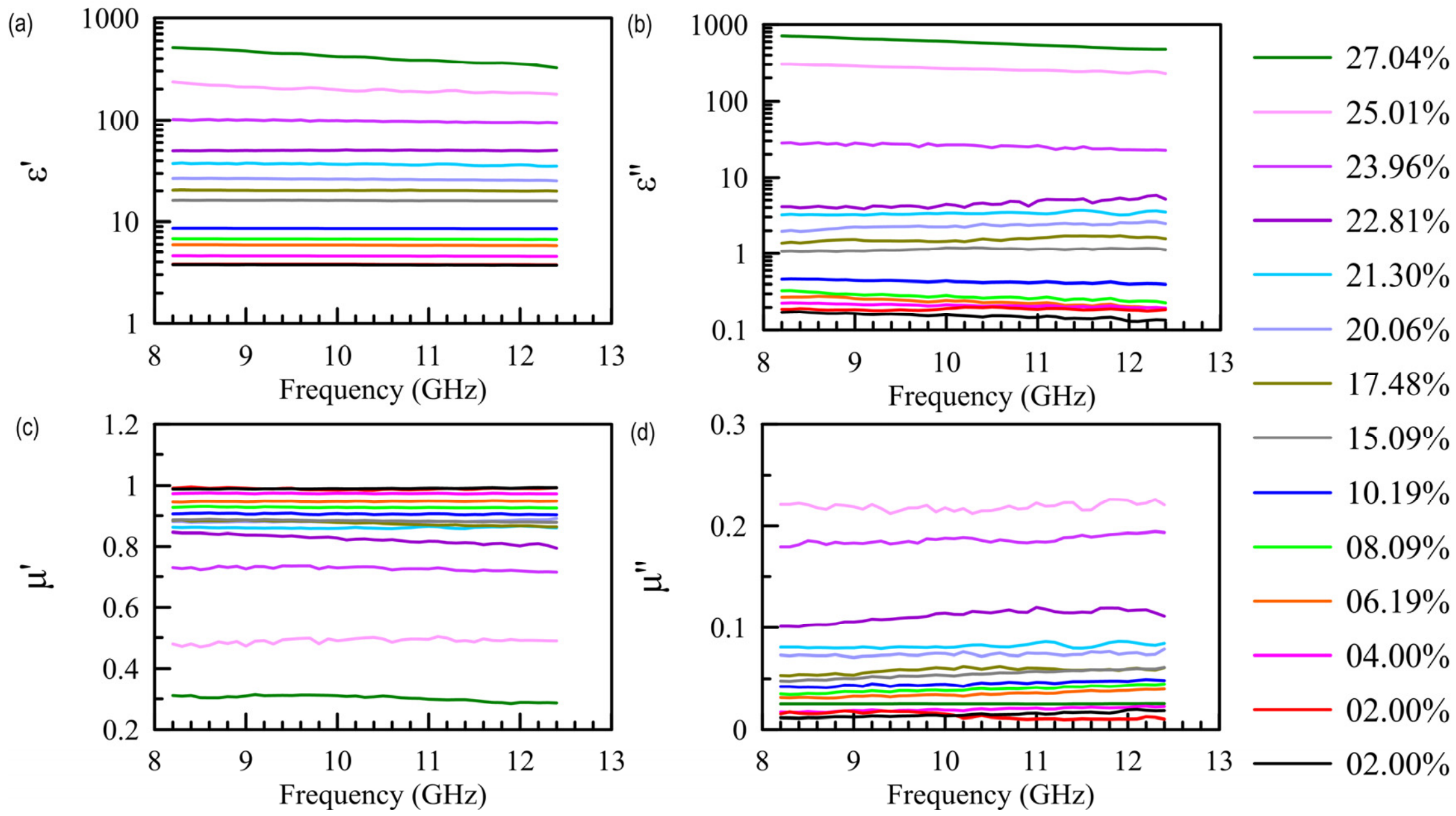

**FIG. 2.** The measured permittivity (a) real part and (b) imaginary part, as well as the measured permeability (c) real part and (d) imaginary part versus frequency for various volume fraction of the silver-epoxy nano-composites from 2.00% to 27.04%. Those complex permittivity and permeability are extracted from the measured scattering parameters based on Eqs. (3) and (4).

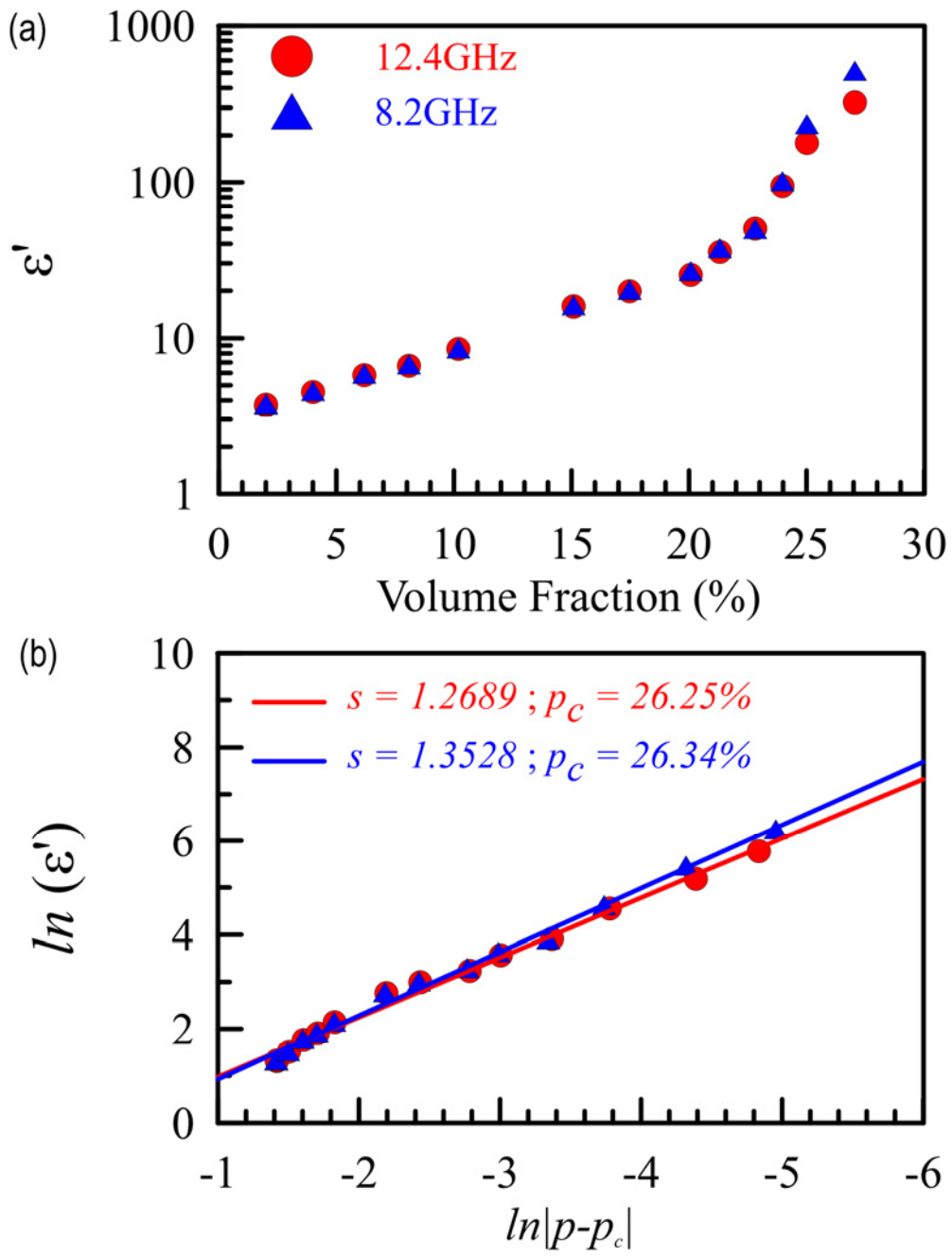

**FIG. 3.** (a) The measured real permittivity at 8.4 GHz (solid triangles) and 12.4 GHz (solid dots) on a logarithmic scale. (b) The fitting result to the percolation theory in Eq. (1a).

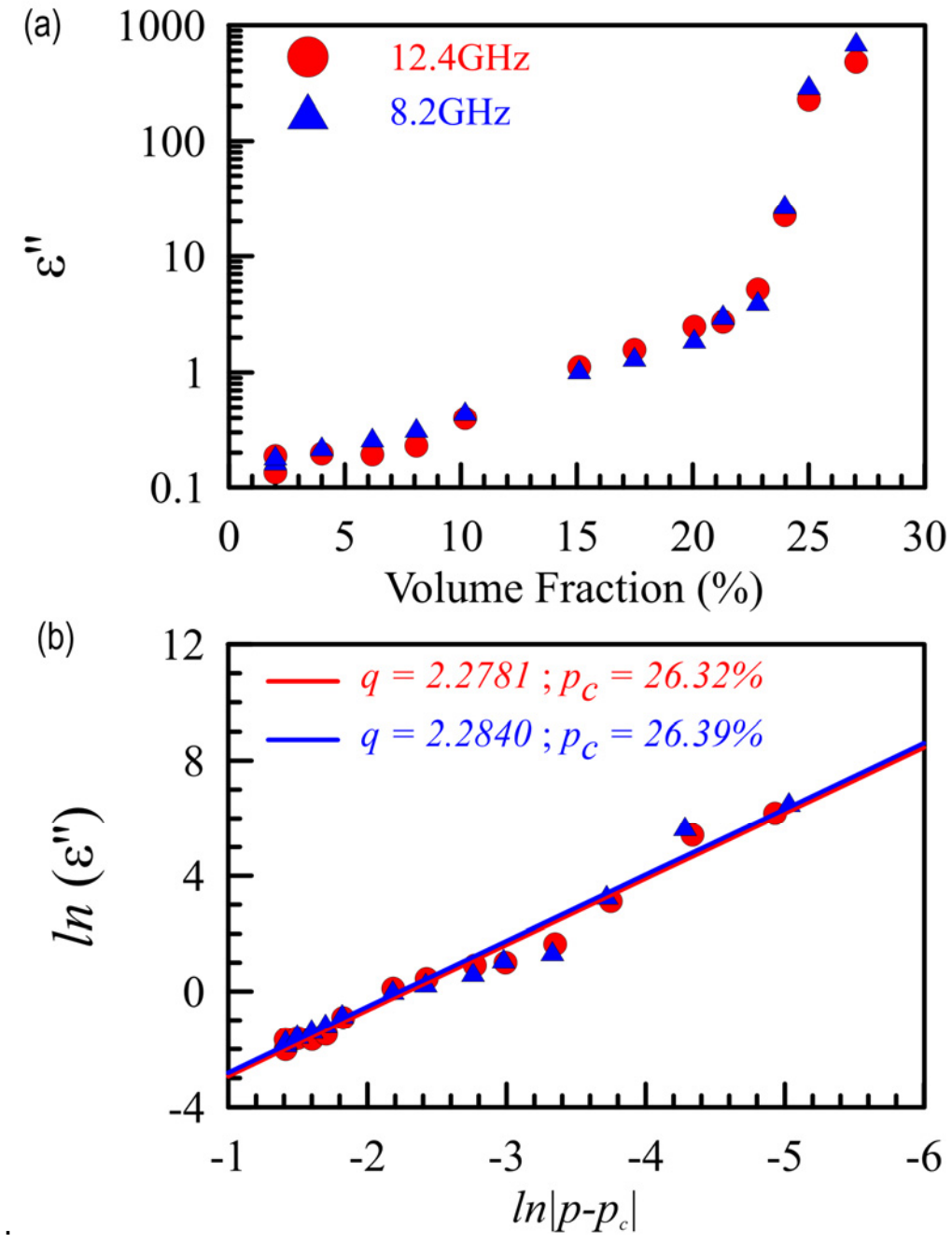

**FIG. 4.** (a) The result of the imaginary permittivity at 8.4 GHz (solid triangles) and 12.4 GHz (solid dots) on a logarithmic scale. (b) The fitting result to the percolation theory in Eq. (1a).

which lie between 25.01% and 27.04%. From the fitting results, the critical exponents $s$ are 1.3528 for 8.2 GHz and 1.2689 for 12.4 GHz. The theoretical value of $s$ is 0.7~1.0 for three-dimensional problems and 1.1~1.3 for two-dimensional problems. The result in this work is closer to the upper limit in two-dimensional cases. The dimension of a WR90 waveguide is 22.86 mm by 10.16 mm, and the thickness chosen was 2-3mm. The shape of the samples resembles a flat slab. This might explain why the critical exponent is high.

Fig. 4(a) shows the imaginary permittivity $\varepsilon''$ at 8.2 GHz (blue triangles) and 12.4 GHz (red dots). It has a similar trend as the real part $\varepsilon'$. A gradual increase occurs below 10%. A more pronounced rise was observed from 15% to 20% and a significant leap at 25%. Then the increasing rate slows down when raised to 27.04%, and reaches a maximum value of 716. The fitting results to Eq. 2(a) are given in Fig. 4(b) on a logarithmic scale. The percolation thresholds for the two frequencies are determined to be 26.32% and 26.39%, which also lies in the range where the increasing rate slows down. The critical exponents $q$ are 2.2781 and 2.2840, which is lower than the theoretical value. The critical exponent $q$ is approximately $s + t$; therefore, the 2D value is about 2.2~2.6, which agrees with our finding.

According to Eq. 1(b) and Eq. 2(b) and Figs. 2(a) and 2(b), the complex permittivity is frequency dependence near the percolation threshold. Thus, the fitting result of both $\varepsilon'$ and $\varepsilon''$ to the frequency is exhibited in Fig 5. The exponents for the volume fraction of 25.01% are 0.5472 (real) and 0.6561 (imaginary), and for 27.04% are 1.0380 (real) and 1.0675 (imaginary), which is highly consistence with Eq. 1(b) and Eq. 2(b). However, the theoretical value of the frequency dependence critical exponent should be somewhere between 0 and 1[32]. The one from 27.04% is slightly higher than the theoretical value.

The permeability results are depicted in Fig. 6(a), and for ease of discussion, the negative magnetic susceptibility is also illustrated in Fig. 6(b). The difference between the two frequencies is small. At low volume fractions (<10%), the relationship between the negative susceptibility and volume fraction is quite linear, as in Fig. 6(c). Most particles remain isolated within this range, and the cluster size remains small, making the percolation effect less apparent. Instead, the scattering model (or the EMT) under the long-range approximation provides a more suitable explanation for this phenomenon.

However, as the volume fraction surpasses 15%, the impact of percolation becomes increasingly dominating. The negative magnetic susceptibility increases slowly from about 0.1 to around 0.2 when the volume fraction is raised from 15% to 24%. Then, it undergoes a remarkable surge of about 0.5 at 25%, reaching a peak of approximately 0.7 at 27%.

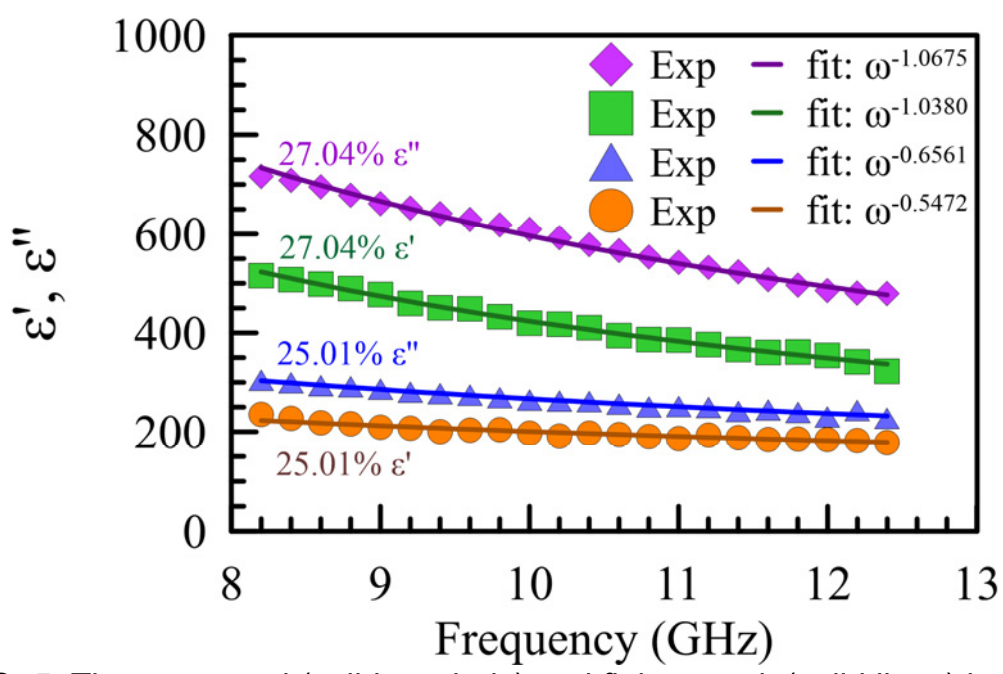

**FIG. 5.** The measured (solid symbols) and fitting result (solid lines) base on equation (1b) of the frequency dependence exponent for both the real and imaginary permittivity in 25.01% and 27.04%. (27.04% $\varepsilon''$ is in purple diamonds, and $\varepsilon'$ in green squares. 25.01% $\varepsilon''$ is in blue triangles, and $\varepsilon'$ in orange dots.)

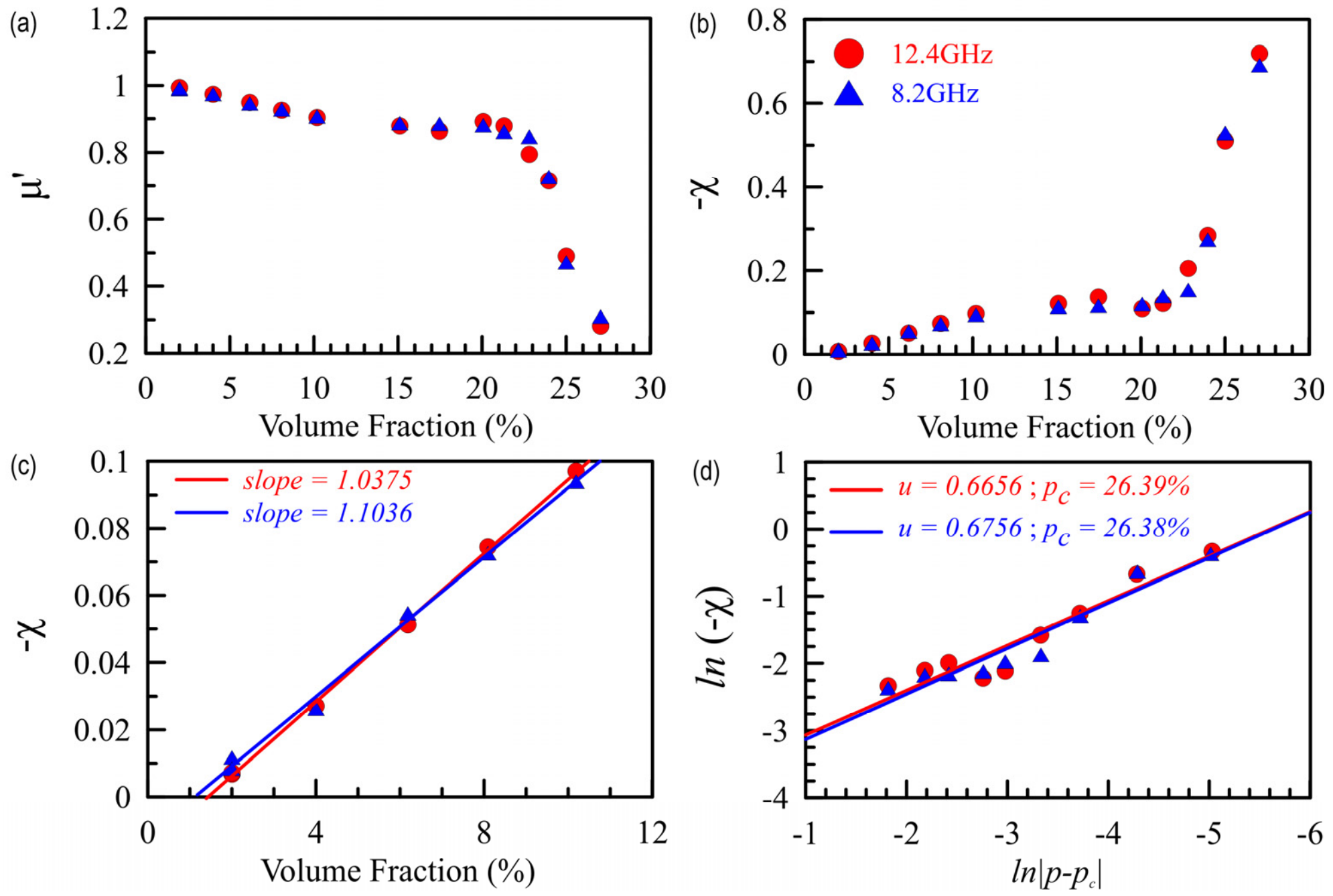

**FIG. 6.** The measured result (solid symbols) of the (a) real permeability and (b) negative magnetic susceptibility with respect to the volume fraction at 8.2GHz (blue) and 12.4GHz (red), and the fitting result (solid lines) in (c) the scattering region and (d) the percolation region.

The fitting curve of the low volume fraction range is given in Fig. 6(c). It should follow the relation [28, 30]

$$-\chi_b = -N\frac{\vec{m}}{\vec{H}_{inc}} = C \cdot p, \qquad (5)$$

where $N$ is the number density of the silver nano-particle. The magnetic dipole to magnetic field relation is $\vec{m} \sim -2S^2\vec{H}_{inc}/P$, where $S$ is the area projected perpendicular to the field, and $P$ is its perimeter.

The relation between the volume fraction and the number density is $p_c = V \cdot N$, where $V$ is the volume of the nano-particle. For perfect spheres, $C = 2S^2/PV$ is 1.5. However, the measured slopes are 1.0375 and 1.1036 for the frequencies of 8.2 and 12.4 GHz, respectively. The obtained slopes suggest that the silver nano-particles are not ideal spheres. The value for a cubic shape is 0.5, and for any shape with approximately the same volume, the value lies between that of a sphere and a cube. Therefore, the value for an arbitrarily shaped particle should fall within the range defined by a sphere and a cube, which aligns with our obtained result.

For the vol. fraction near the threshold case, the fitting results are depicted in Fig. 6(d) on a logarithmic scale by Eq. (3). The threshold is determined to be 26.38% and 26.39%, closely aligning with the earlier obtained value. Moreover, the critical exponent $u$ is calculated to be 0.6756 and 0.6656, which is almost twice the value of the theoretical one. However, the theoretical value in 2D is about 0.77 [28, 34], which the result is closer to. The observed permeability's power law behavior in this study exhibits strong consistency with the predictions from Ref. [28].

Fig. 7(a) displays the measured imaginary permeability $\mu''$ for two representative frequencies 8.2 and 12.4 GHz. It grew slowly and linearly when the volume fraction was increased from 2% to 10%. Then, the rate of growth paced down between 15% to 24%. As the volume fraction was further increased, the rate of increment surged swiftly to a maximum of 0.24 at p = 25.01% and dropped to around 0.02 when the threshold was exceeded. The fitting result based on Eq. (4) is given in Fig. 7(b) on a logarithmic scale. The result follows the power law. The threshold is determined to be 26.60% and 25.92%, which slightly differs from the result acquired prior, but still in a reasonable region. The critical exponent $v$ is 0.6987 and 0.6567, which is highly similar to $u$ in 2D [28, 34].

The result supports Bowman and Stroud's prediction for insulator-conductor composite. Otherwise, if we neglect the permeability's contribution, the result deviates, which is given in supplementary material (Fig. SM01). Both the real and imaginary permittivity should monotonousely increase when the vol. fraction approaches to the threshold. However, the imaginary permittivity (at 12.4GHz) increases and decrease multiple times when $p = 17.48\% \sim 27.04\%$.

The complex permittivity of $\mu = 1$ to $\mu \neq 1$ was compared, which is illustrated in the supplementary material (Fig. SM02). The percolation phenomenon is less obvious, especially in the imaginary part. Also, both $\mathrm{Re}[\varepsilon]$ and $\mathrm{Im}[\varepsilon]$ are smaller than the case $\mu \neq 1$. Finally, we can compare the loss tangent ($\tan\delta_\varepsilon$). The loss tangent should also surge near the threshold. The loss tangent is displayed in the supplementary material (Fig. SM03). Loss tangent, $\delta_\varepsilon$, surges when we count permeability's contribution. Our experiment result suggests that permeability can vary in percolative metal-insulator composite near the percolation threshold. When dealing with a metal-insulator composite system, the effect due to permeability should be taken into account.

Based on percolation theory, we conclude that (a) when the vol. fraction approaches the phase transition threshold, the permittivity increases while the permeability drops, as predicted. (b) The effective medium theory's validation seems weakened near the percolation threshold. (c) In this case, the permittivity and permeability are not equal to 1 near the percolation threshold. For permittivity- and permeability-related quantities, such as the refraction index, it will be necessary to include both permittivity and permeability near the percolation threshold. Otherwise, the calculated refraction index could deviate from the experiment result.

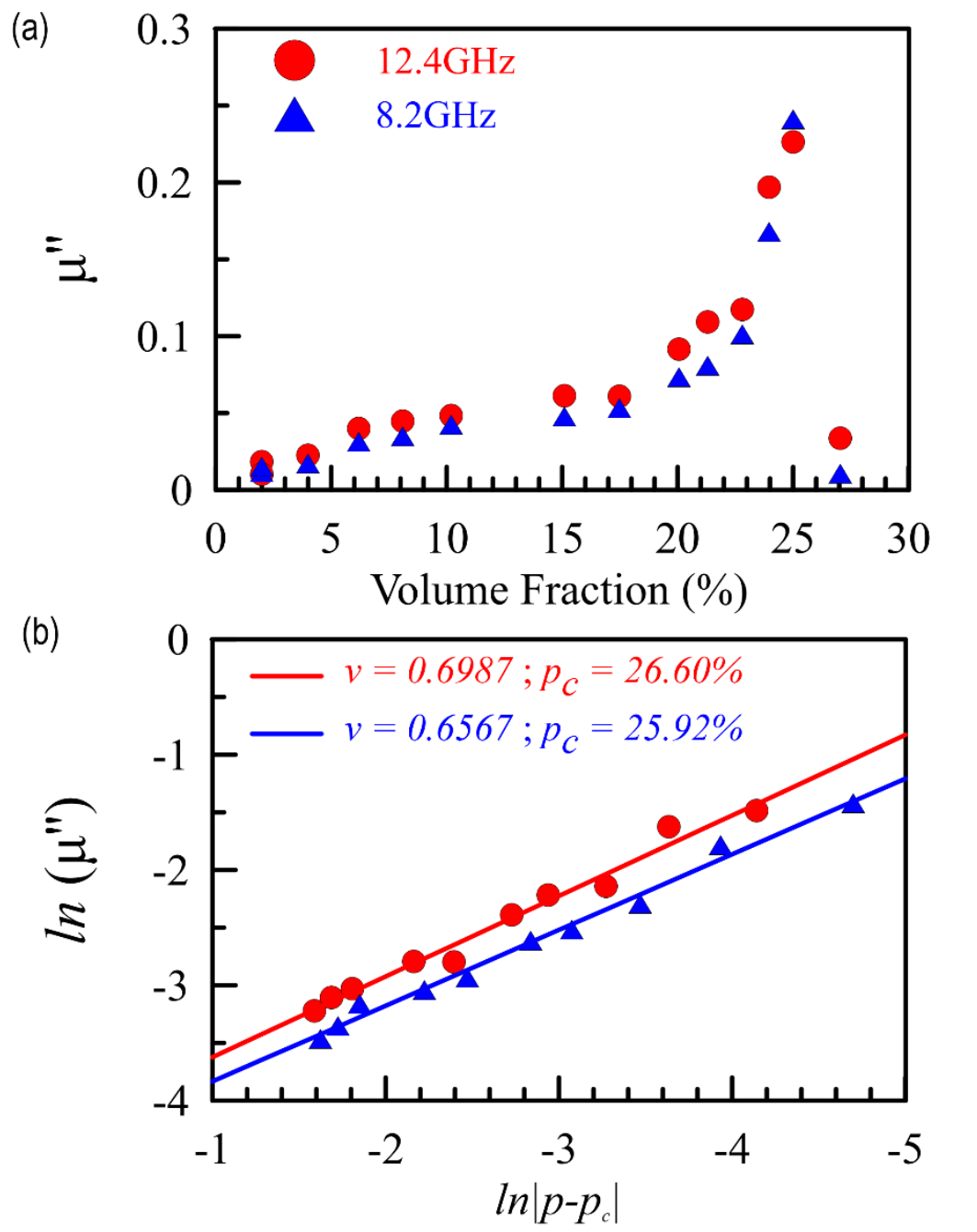

**FIG. 7.** (a) The result of the imaginary permeability at 8.4 GHz (solid triangles) and 12.4 GHz (solid dots) on a logarithmic scale. (b) The fitting result to the percolation theory in Eq. (4).

The percolation theory characterizes composite materials' behavior near its percolation threshold. All EM properties, permittivity, and permeability are power law dependencies. Below the threshold (far away), the composite is in the insulator (dielectric) phase, but there's a phase transition around the threshold that leads to the conduction phase. While there's been extensive research on the transition from insulator to conductor, percolation theory for permeability hasn't received as much attention comparatively. In detail, the identified percolation threshold is approximately 25.92% to 26.60%, which is relatively high [35, 36]. The fitted critical exponents indicate that our system is closer to a 2D percolation than a 3D one, which could be attributed to the cross-section-to-thickness ratio; the percolation threshold of 2D cases is higher than 3Ds [31, 32].

Our findings align closely with Bowman and Stroud's predictions, particularly regarding the power-law behavior of negative susceptibility, as expressed in Eq. (3). Crucially, the results presented here hold validity across a broad bandwidth (8.2-12.4GHz). This also implies that the insulator-conductor composite has a non-one permeability in THz region, since the percolation threshold is not frequency dependence. The authors anticipate that this work will draw attention to the importance of percolation theory on complex permittivity and permeability.

## ACKNOWLEDGMENTS

This work was supported by the National Science and Technology Council under the contract No. NSTC 113-2112-M-007-023-MY3. The authors are also grateful to Taiwan Semiconductor Research Institute, TSRI for the measurement support.

## AUTHOR DECLARATIONS

### Conflict of Interest

The authors have no conflicts to disclose.

### Author Contributions

**Bo-Wei Tseng**: Data curation (lead); Formal analysis (lead); Validation (lead); Formal analysis (lead); Writing – original draft (lead); Writing – review & editing (equal).

**Tsun-Hsu Chang**: Data curation (supporting); Formal analysis (supporting). Project administration (lead); Supervision (lead); Writing – review & editing (equal). Contributions are subject to change.

## DATA AVAILABILITY

The data supporting this study's findings are available from the corresponding author upon reasonable request.